\documentclass[aps,prl,twocolumn,showpacs,superscriptaddress,nofootinbib,nobibnotes]{revtex4-1}

\usepackage{breqn}
\usepackage{graphicx}
\usepackage{dcolumn}   
\usepackage{bm}   
\usepackage{amssymb}
\usepackage{soul}
\usepackage{gensymb}

\usepackage[breaklinks=true,colorlinks=true,linkcolor=blue,urlcolor=blue,citecolor=blue]{hyperref}

\usepackage{float}

\begin{document}

\title{New high-confinement regime with fast ions in the core of fusion plasmas}

\author{A.~Di Siena} 
\affiliation{The University of Texas at Austin 201 E 24th St 78712 Austin Texas USA}
\author{R.~Bilato}
\affiliation{Max Planck Institute for Plasma Physics Boltzmannstr 2 85748 Garching Germany}
\author{T.~G\"orler}
\affiliation{Max Planck Institute for Plasma Physics Boltzmannstr 2 85748 Garching Germany}
\author{A.~Ba\~n\'on~Navarro}
\affiliation{Max Planck Institute for Plasma Physics Boltzmannstr 2 85748 Garching Germany}
\author{E.~Poli}
\affiliation{Max Planck Institute for Plasma Physics Boltzmannstr 2 85748 Garching Germany}
\author{V.~Bobkov}
\affiliation{Max Planck Institute for Plasma Physics Boltzmannstr 2 85748 Garching Germany}
\author{D.~Jarema}
\affiliation{Max Planck Institute for Plasma Physics Boltzmannstr 2 85748 Garching Germany}
\author{E.~Fable}
\affiliation{Max Planck Institute for Plasma Physics Boltzmannstr 2 85748 Garching Germany}
\author{C.~Angioni}
\affiliation{Max Planck Institute for Plasma Physics Boltzmannstr 2 85748 Garching Germany}
\author{Ye.~O.~Kazakov}
\affiliation{Laboratory for Plasma Physics LPP-ERM/KMS TEC Partner 1000 Brussels Belgium}
\author{R.~Ochoukov}
\affiliation{Max Planck Institute for Plasma Physics Boltzmannstr 2 85748 Garching Germany}
\author{P.~Schneider}
\affiliation{Max Planck Institute for Plasma Physics Boltzmannstr 2 85748 Garching Germany}
\author{M.~Weiland}
\affiliation{Max Planck Institute for Plasma Physics Boltzmannstr 2 85748 Garching Germany}
\author{F.~Jenko} 
\affiliation{Max Planck Institute for Plasma Physics Boltzmannstr 2 85748 Garching Germany}
\author{the ASDEX Upgrade Team}
\thanks{See the author list of H. Meyer et al., Nucl. Fusion 59 (2019), 112014}

\begin{abstract}

The key result of the present work is the theoretical prediction and observation of the formation of a new type of transport barrier in fusion plasmas, called F-ATB (fast ion-induced anomalous transport barrier). As demonstrated through state-of-the-art global electrostatic and electromagnetic simulations, the F-ATB is characterized by a full suppression of the turbulent transport - caused by strongly sheared, axisymmetric $E \times B$ flows - and an increase of the neoclassical counterpart, albeit keeping the overall fluxes at significantly reduced levels. The trigger mechanism is shown to be a mainly electrostatic resonant interaction between supra-thermal particles, generated via ion-cyclotron-resonance heating, and plasma micro-turbulence. These findings are obtained by realistic simulations of the ASDEX Upgrade discharge $\#36637$ - properly designed to maximized the beneficial role of the wave-particle resonance interaction - which exhibits the expected properties of improved confinement produced by energetic particles. 

\end{abstract}

\pacs{52.65.y,52.35.Mw,52.35.Ra}

\maketitle

{\em Introduction.} The performance of present-day and future fusion devices is largely determined by turbulent transport. Turbulence is inevitably driven by gradients in the plasma pressure profiles above a critical threshold, and it is the primary source of energy and particle losses. Therefore, a long-standing challenge in magnetic fusion research is the identification of mechanisms able to control turbulence with the ultimate goal of improving reactor performances.

In this context, a fascinating result is the possibility of forming a narrow core region of reduced turbulent transport, commonly called internal transport barrier (ITB) \cite{Conway_PRL_2000,Connor_NF_2004}. ITBs are characterized by large poloidal shear flows localized at the radial boundaries of the barrier, acting on turbulent transport through a combination of linear and nonlinear effects \cite{Burrell_PoP_1997,Burrell_PoP_2020}. The formation of an ITB has been observed in both tokamaks (DIII-D \cite{Rice_PoP_1996}, ASDEX Upgrade \cite{Gruber_PPCF_2000,Tardini_NF_2007}, JT60-U \cite{Fujita_PRL_1997}, TFTR \cite{Levinton_PRL_1995}, Tore Supra \cite{Litaudon_PPCF_1999}, Alcator C-Mod \cite{Ernst_PoP_2004, Zhurovich_NF_2007}, JET \cite{Gormezano_PPCF_1999}) and stellarators (LHD \cite{Ida_PRL_2003}, CHS \cite{Fujisawa_PRL_1999}, TJ-II \cite{Castejon_NF_2002} and W7-AS \cite{Stroth_PRL_2001}). In these experiments, a ubiquitous observation - associated with the ITB formation - is the increase of the fusion triple product. The main mechanisms proposed to explain the formation of ITBs in tokamaks ware (i) the impact of large-scale electromagnetic activity at low-order rational surfaces in the safety factor \cite{Joffrin_NF_2003}, often associated to reversed-shear configurations \cite{Joffrin_NF_2002,Joffrin_PPCF_2002,Joffrin_NF_2003} and (ii) electromagnetic supra-thermal ion effects \cite{Romanelli_PPCF_2010}.

In this Letter, we present a novel type of transport barrier induced by fast ions created with ion-cyclotron-resonance-heating (ICRH) called F-ATB (fast ion-induced anomalous transport barrier). The F-ATB is characterized by a full suppression of the turbulent heat transport and an increase in the neoclassical counterpart - driven by the large pressure gradients of the supra-thermal particles. The resulting overall turbulence levels are reduced and the transition from anomalous to neoclassical transport (within the barrier) leads to a substantial de-stiffening of the bulk profiles. In contrast to ITBs reported in literature, this new type of transport barrier is stable and easily controllable with the ion-cyclotron-resonance-heating (ICRH).

\begin{figure*}
\begin{center}
\includegraphics[scale=0.27]{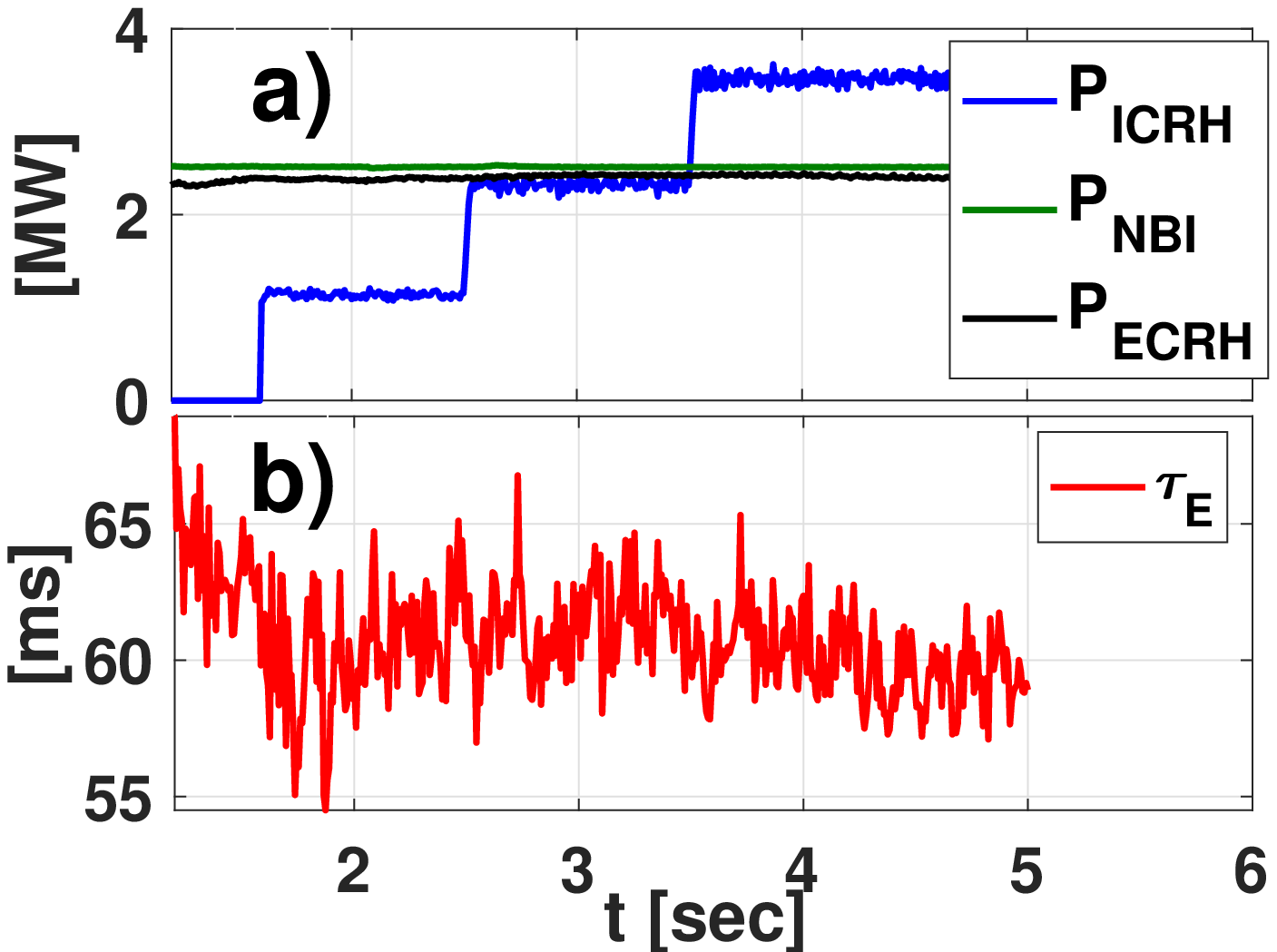}\includegraphics[scale=0.27]{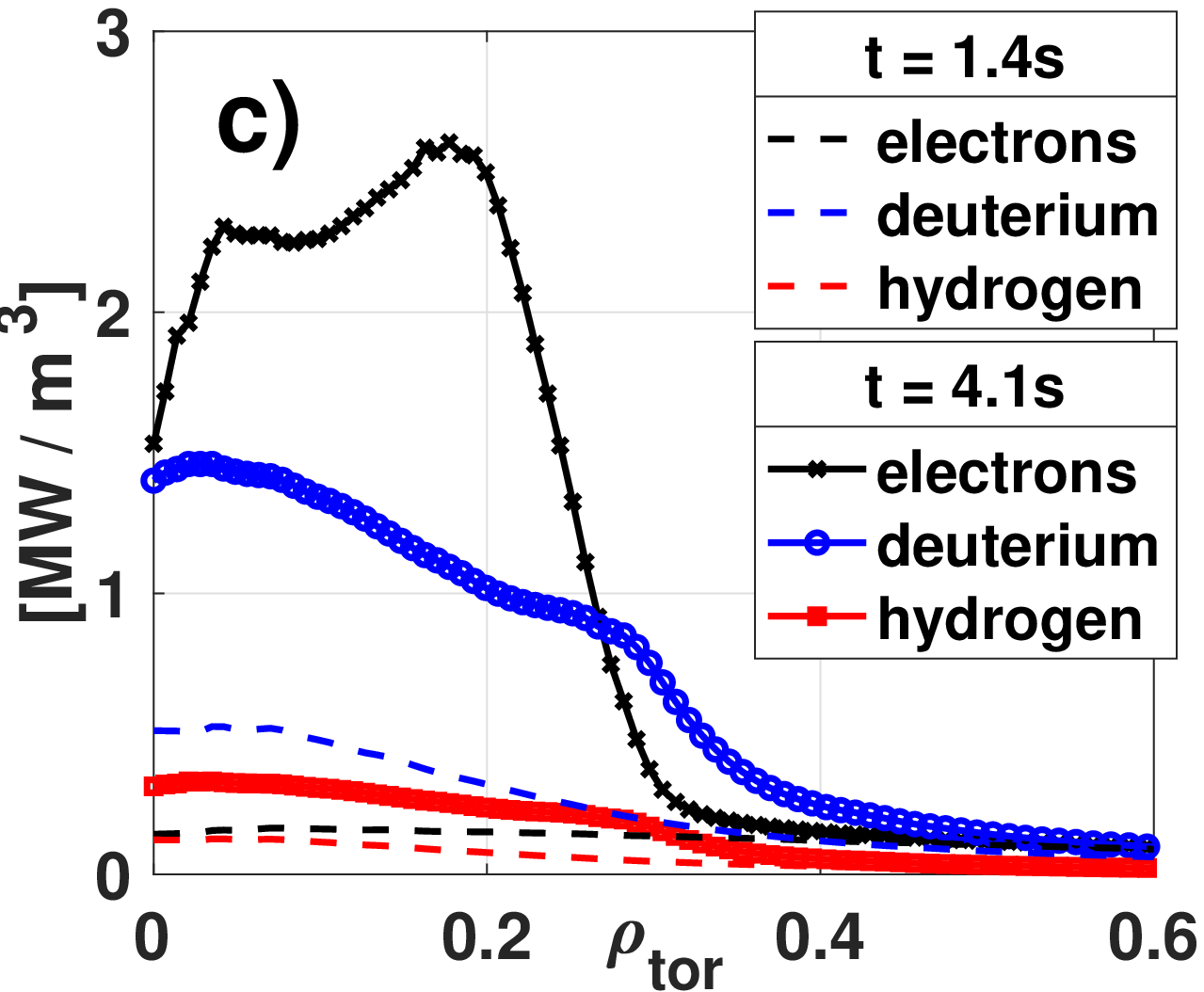}\includegraphics[scale=0.27]{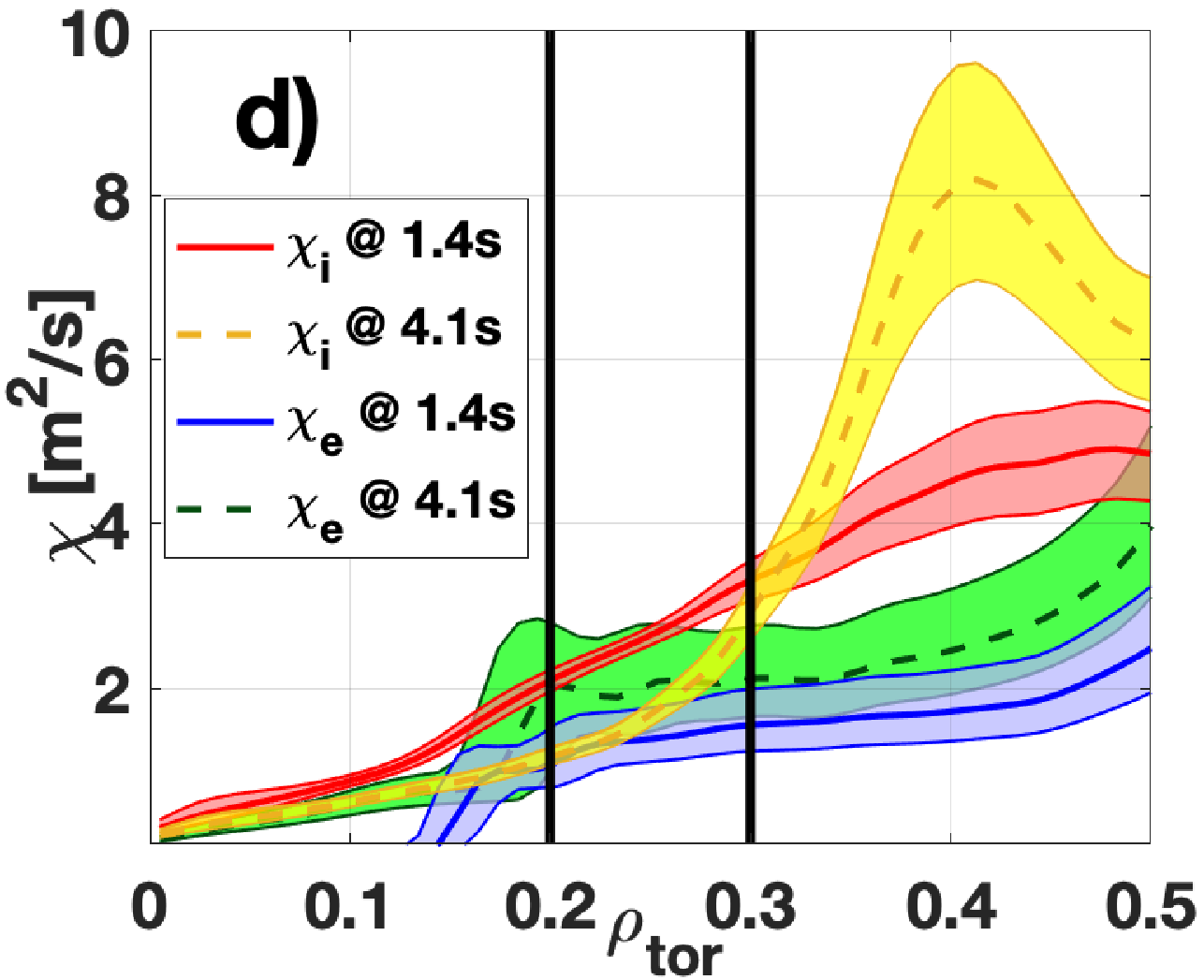}\includegraphics[scale=0.27]{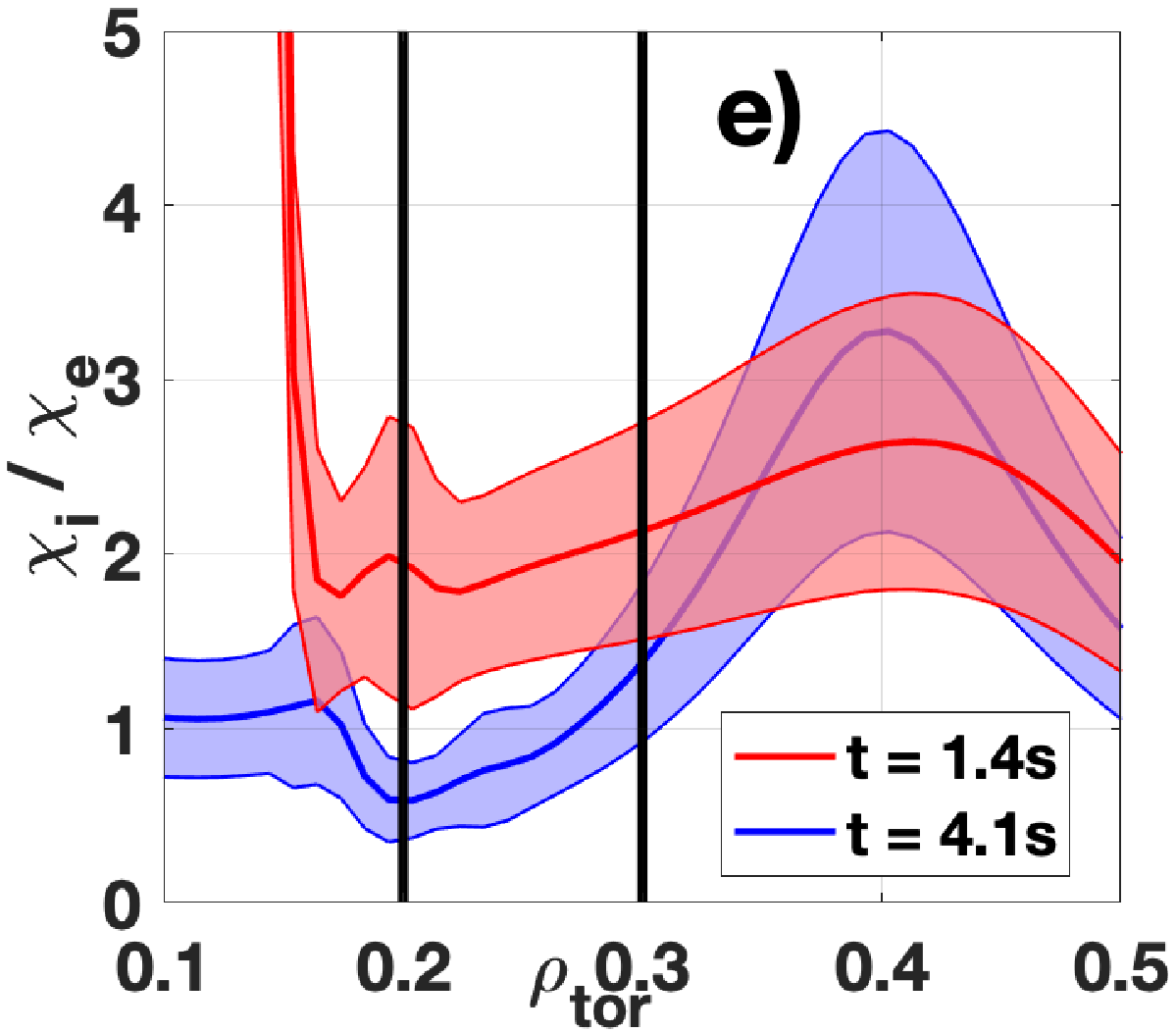}\includegraphics[scale=0.27]{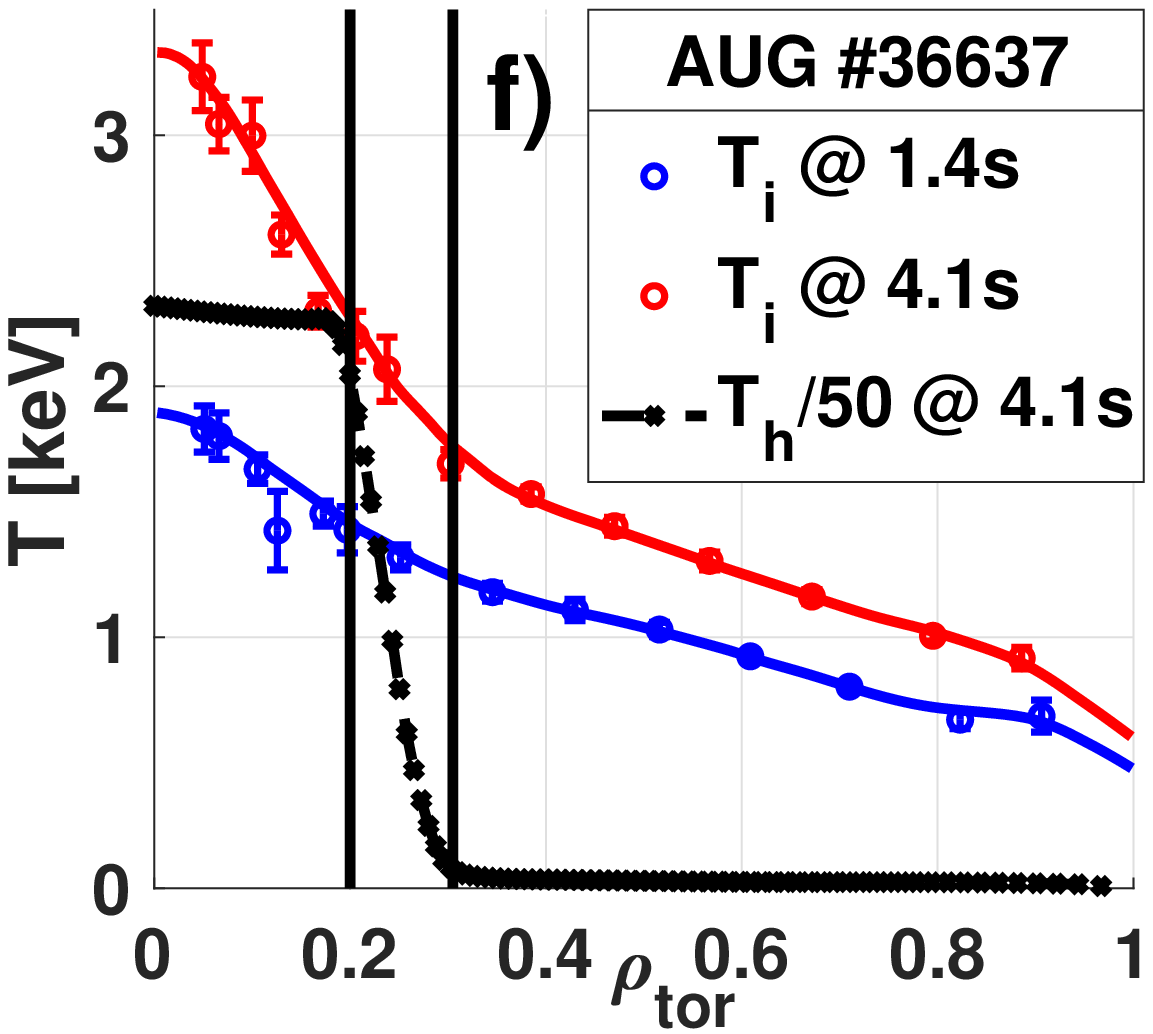}
\par\end{center}
\caption{Time history of a) injected power and b) confinement time for the ASDEX Upgrade discharge $\#36637$; c) Power repartition (NBI + ICRF) among the different plasma species at $t=1.4$s and $t=4.1$s as computed by TORIC/SSFPQL by retaining the effect of collisions; d) Ion and electron thermal conductivities ($\chi_i$ and $\chi_e$, expressed in SI units) computed by ASTRA at $t=1.4$s and $t=4.1$s; e) Ratio between $\chi_i/\chi_e$; f) Main ion temperature profiles $(T_i)$ at $t=1.4$s (blue) and $t=4.1$s (red) and equivalent fast-ion temperature of distribution function of the hydrogen minority ($T_h$ re-scaled by a factor of 50), computed by TORIC-SSFPQL at $t=4.1$s (black). The vertical black lines denote the region where the logarithmic temperature gradients deviate the most.}
\label{fig:fig1}
\end{figure*}

The F-ATB trigger mechanism is identified as the (mainly electrostatic) resonant interaction between ion-driven turbulence and supra-thermal particles, recently discovered via analytic calculations and corroborated by (flux-tube) numerical simulations \cite{DiSiena_NF2018,DiSiena_PoP2019,W7X}. The impact of this resonance mechanism on plasma turbulence is determined mainly by the local values of the fast particle temperature, density, and their gradients \cite{DiSiena_NF2018}. Therefore, stabilizing and destabilizing energetic particle effects on plasma turbulence might occur at different radial positions - depending on the local fast particle parameters - thus strongly affecting the bulk ion energy fluxes. This characteristic radial dependence has been carefully optimized with gyrokinetic and ICRH-full-wave simulations, allowing us to identify the main experimental actuators to maximize this resonant mechanism. This effort guided us in designing and performing a proof-of-principle ASDEX Upgrade discharge showing features of transport reduction and the formation of a central region of improved confinement produced by energetic particles. 
On this plasma discharge we then performed state-of-the-art global electromagnetic GENE \cite{Jenko_PoP2000,Goerler_JCP2011} simulations, demonstrating - for the first time - an anomalous transport barrier in these theory-guided plasma conditions, thus providing the numerical confirmation of the existence of this new type of ITB.

{\em Experimental setup.} The ASDEX Upgrade H-mode discharge $\# 36637$ was the result of a careful optimization procedure based on theoretical predictions \cite{DiSiena_NF2018,DiSiena_PoP2019}. A deuterium plasma is heated with $2.5$MW of neutral-beam-injected (NBI) and $2.5$MW of electron-cyclotron-resonance-heating (ECRH) power. An additional ion-cyclotron-resonance-heating is applied with four steady-state phases at constant power (Fig.~\ref{fig:fig1}a). 
The ICRH is deposited on-axis ($\rho_{\rm tor} = 0$) on a large hydrogen minority density at a roughly constant concentration of $n_H/n_e \approx 0.11$, inferred from neutral-particle-analyzer measurements, resulting in $T_{h}(0) \sim 110$keV (Fig.~\ref{fig:fig1}f). The magnetic field is $B_0 = 2.38 $T, the plasma current $I_p = 0.8$MA and the mid-radius electron density, temperature and gyroradius respectively, $n_e = 5.2 \cdot 10^{19}{\rm m^{-3}}$, $T_e = 1.8$keV and $\rho^* = \rho/a = 1/385$. Despite most of the external NBI+ICRF power being absorbed by electrons (Fig.~\ref{fig:fig1}c), only the main ion temperature profile (measured by IDI \cite{Fischer_FST_2020}) develops a substantial peaking, up to $~80\%$ (from $T_i(0) = 1.9$keV to $T_i(0) = 3.4$keV) for $P_{\rm ICRH} = 3.5$MW, while the overall core electron temperature profile is up-shifted by $20\%$ whilst keeping a similar shape and the density profile remains almost unaffected.

No significant degradation of the energy confinement is observed during the ramp-up of the ICRH power (Fig.~\ref{fig:fig1}b). Interestingly, the power balance computed by ASTRA \cite{Fable_PPCF_2013} (Fig.~\ref{fig:fig1}d) shows that (i) the ion conductivity at $t = 4.1$s (compared to $t = 1.4$s) is reduced up to $\sim 50\%$ at $\rho_{\rm tor} = 0.2$ despite an increase of $\sim40\%$ of the auxiliary heating. It increases for $\rho_{\rm tor} > 0.3$; (ii) the electron conductivity measured at $t = 1.4$s and $t = 4.1$s increases, leading to a further reduction in the ratio $\chi_i / \chi_e$ (Fig.~\ref{fig:fig1}e). These findings suggest a substantial ion-scale turbulence suppression. Additional dedicated experiments on ASDEX Upgrade are presently in progress~\cite{Bilato_2020}.

{\em Numerical setup.} The ICRH fast particles role in improving the plasma confinement is investigated with the global gyrokinetic code GENE \cite{Jenko_PoP2000,Goerler_JCP2011}. The experimental plasma equilibrium is reconstructed by TRACER-EFIT \cite{Xanthopoulos_PoP_2009} and the bulk profiles extracted from IDA \cite{Fischer_FSC_2010}. Kinetic electrons with realistic ion-to-electron mass ratio are retained and collisions modeled with a linearized Landau operator with energy and momentum conserving terms \cite{Crandall_CPC_2020}. A local Maxwellian is employed to model the thermal species backgrounds, while a bi-Maxwellian distribution \cite{DiSiena_2016,DiSiena_PoP_2018,DiSiena_NF_2018_egam} describes the hydrogen minority. The parallel and perpendicular temperature profiles are consistently computed with the full-wave TORIC code interfaced with the SSFPQL Fokker–Planck solver \cite{Brambilla_PPCF1989,Bilato_2011}. Electromagnetic fluctuations are retained throughout this Letter (unless stated otherwise) and a realistic profile for the electron kinetic-to-magnetic pressure ratio $\beta_e$ is considered, namely $\beta_e = 0.8\%$ on-axis. A Krook-type operator on both heat and particles is employed to keep the kinetic profiles - on average - fixed at the initial ones. Dirichlet boundary conditions are applied on the perturbed quantities. The radial domain of the GENE simulations covers the region $\rho_{\rm tor} = [0.05,0.55]$, where the thermal ion profile and conductivities are mostly affected by the presence of energetic particles. An optimized radially dependent velocity space grid \cite{Jarema_CPC_2016} is employed to reduce the resolution requirements needed to capture the sharp changes in the energetic particle temperature profile. Numerical convergence, over the grid resolution and different velocity space grids, has been carefully checked.

\begin{figure}
\begin{center}
\includegraphics[scale=0.40]{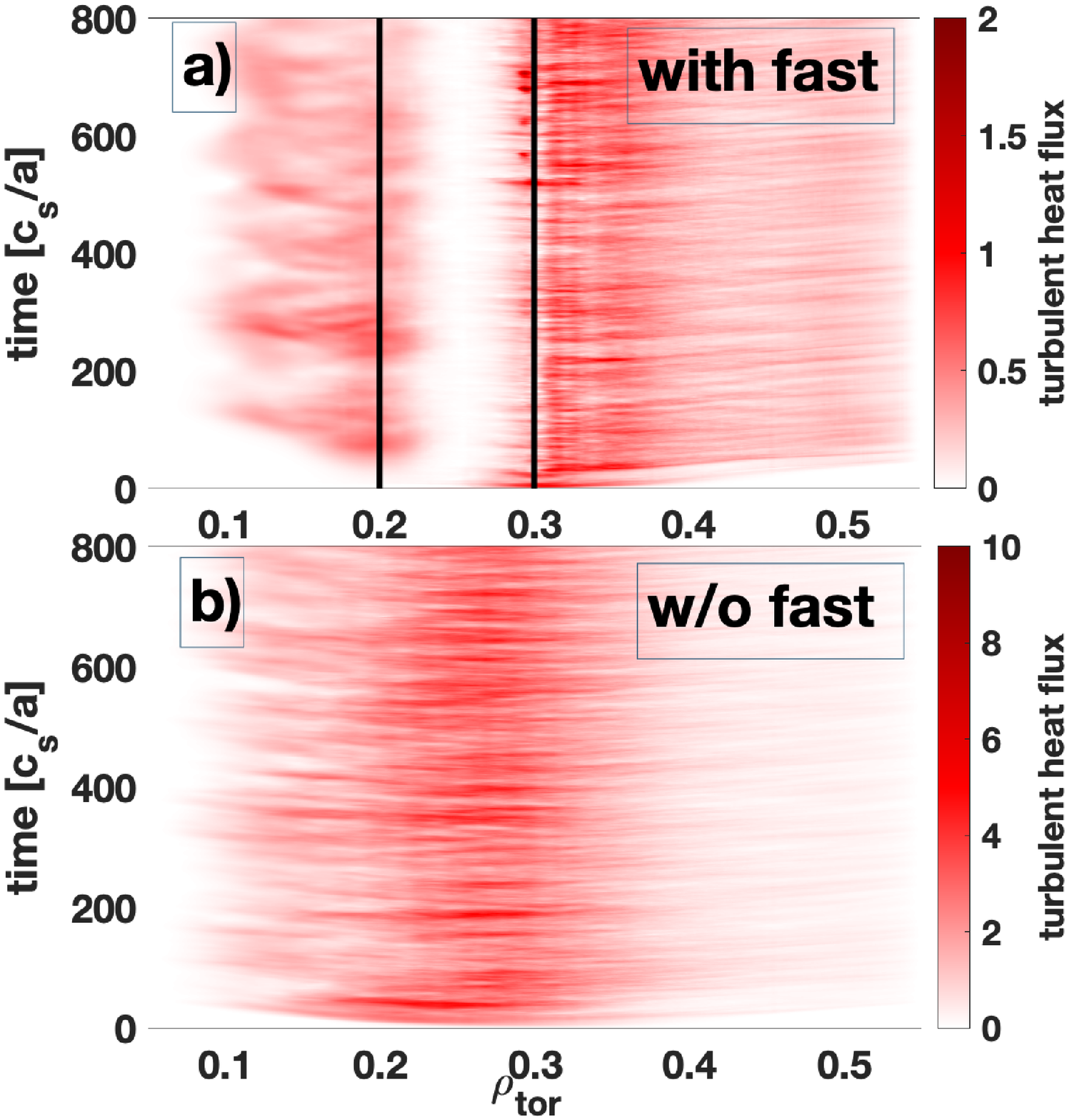}
\includegraphics[scale=0.27]{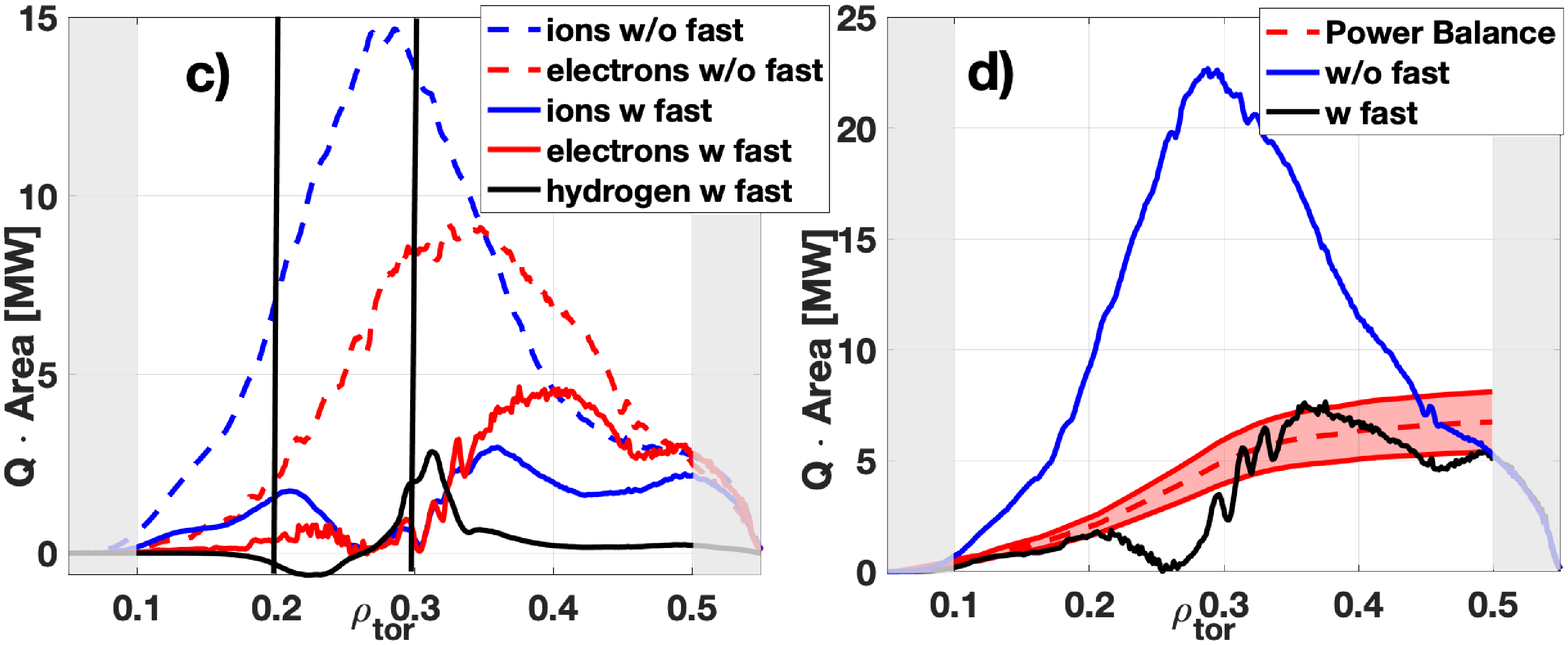}
\par\end{center}
\caption{Time evolution of the radial profile of the total ion heat flux (thermal ions + hydrogen) in gyroBohm units. The same magnetic equilibrium and kinetic bulk profiles are employed in the simulations with (a) and without (b) the supra-thermal ions. Radial profile of the (c) different species and (d) total (thermal ions + electrons + hydrogen) heat flux in MW averaged over $t [c_s / a] = [400 - 600]$. The red line in figure d) represents the volume integral of the injected sources computed by ASTRA. The vertical black lines denote the radial position of the F-ATB, while the shaded-gray areas, the buffer zones.}
\label{fig:fig2}
\end{figure}

{\em Global turbulence results.} We begin by showing, in Fig.~\ref{fig:fig2}, the time evolution of the radial profile of the total ion (deuterium + hydrogen) flux-surface-averaged heat flux for the simulations with (Fig.~\ref{fig:fig2}a) and without (Fig.~\ref{fig:fig2}b) energetic particles. 
Turbulent avalanches propagate throughout the radial domain when only thermal particles are retained. In the presence of fast particles, these structures break in the region $\rho_{\rm tor} = [0.2, 0.3]$ - where the temperature profile of the bulk ion peaks. Within this radial domain, a full suppression of the heat transport is observed, leading to the formation of the F-ATB. Interestingly, the turbulence reduction extends to $\rho_{\rm tor} \sim 0.1$ (Fig.~\ref{fig:fig2}c-d). These findings are consistent with the modifications in the ion conductivity due to the ramp-up of the ICRF power (Fig.~\ref{fig:fig1}d), where $\chi_i$ drops in the region $\rho_{\rm tor} = [0.1, 0.3]$. The heat flux contribution of each plasma species is shown in Fig.~\ref{fig:fig2}c. In the absence of energetic ions, the overall heat flux is largely over-predicted as compared to the volume integral of the injected heating sources computed by ASTRA. A quantitative agreement is obtained outside the barrier only by including the hydrogen minority and electromagnetic fluctuations (Fig.~\ref{fig:fig2}d), thus revealing that the numerical (gradient-driven) GENE setup is already close to the (unfeasible) flux-driven solution. The missing flux within the barrier is likely provided by the neoclassical component (Fig.~\ref{fig:fig4}a). It is worth mentioning that the F-ATB formation is not an artefact of the specific plasma profiles employed but a pervasive observation in our GENE simulations that include the hydrogen minority with sufficiently peaked temperature profiles.

A peculiarity of the F-ATB is the spontaneous self-regulation of persistent, localized shearing layers in correspondence to the radial boundaries of the barrier. 
\begin{figure}
\begin{center}
\includegraphics[scale=0.40]{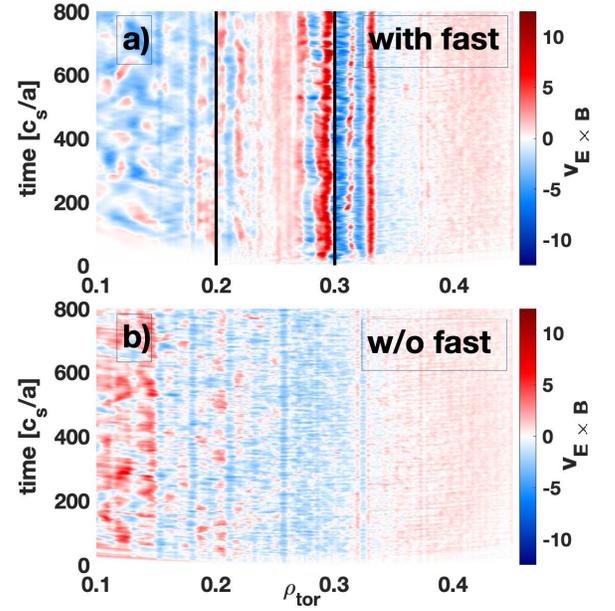}
\par\end{center}
\caption{Time evolution of the radial profile of $v_{E\times B}$ obtained with (a) or without (b) hydrogen minority. The vertical black lines denote the radial position of the F-ATB.}
\label{fig:fig3}
\end{figure}
As observed by looking at the radial profile evolution of the surface-averaged $E \times B$ velocity ($v_{E \times B} = \partial_{\rho_{\rm tor}} \phi_r/\rho_{\rm tor}B_0$, where $\phi_r$ represents the flux-surface-averaged electrostatic potential) of Fig.~\ref{fig:fig3}, an oscillatory pattern develops around $\rho_{\rm tor} = 0.3$ which is reminiscent of the results found in Refs.~\cite{Waltz_PoP_2006,Dif-Pradalier_PRL_2015}. These characteristic features disappear when the hydrogen minority is excluded from the simulations. 

\begin{figure}
\begin{center}
\includegraphics[scale=0.27]{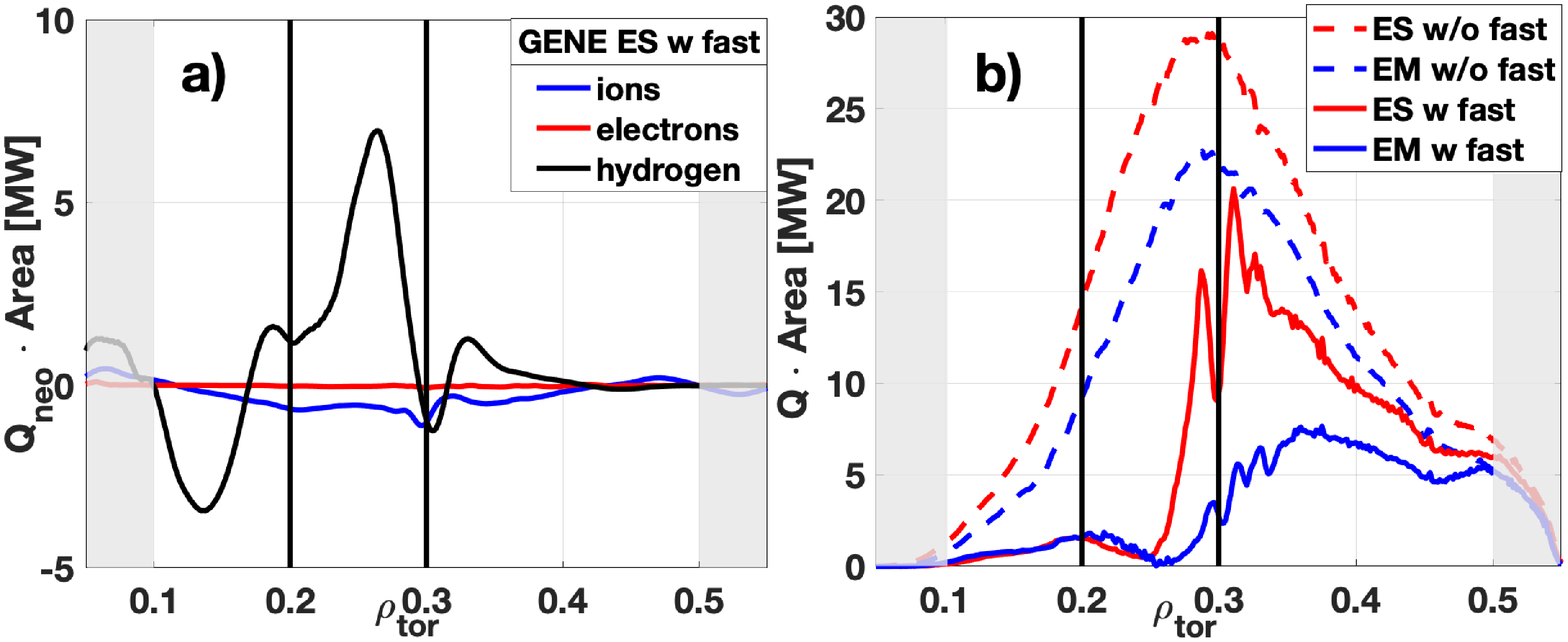}
\par\end{center}
\caption{Radial profile of the a) electrostatic neoclassical heat flux contribution of each species in MW averaged over $t [c_s / a] = [1250 - 1350]$ and b) total heat flux (thermal ions + electrons + hydrogen) in MW averaged over $t [c_s / a] = [400 - 600]$ . The blue lines in (b) represent the electromagnetic results while the red ones the electrostatic ones with (continuous) and without (dotted) the hydrogen minority. The vertical black lines denote the radial position of the F-ATB, while the shaded-gray areas, the buffer zones.}
\label{fig:fig4}
\end{figure}
As the turbulent heat flux drops in the F-ATB region, a corresponding increase of the neoclassical transport to the turbulent levels - dominated by the hydrogen minority contribution - is observed. This is shown in Fig.~\ref{fig:fig4}a, where the time-averaged radial profile of the neoclassical heat flux is illustrated for the electrostatic GENE simulation, retaining both turbulence and neoclassical effects \cite{Oberparleiter_PoP_2016}. A similar behaviour of the neoclassical fluxes is expected in the more computationally expensive electromagnetic setup, for which a full simulation up to the ion-ion collisional time would be unfeasible.

\begin{figure*}
\begin{center}
\includegraphics[scale=0.35]{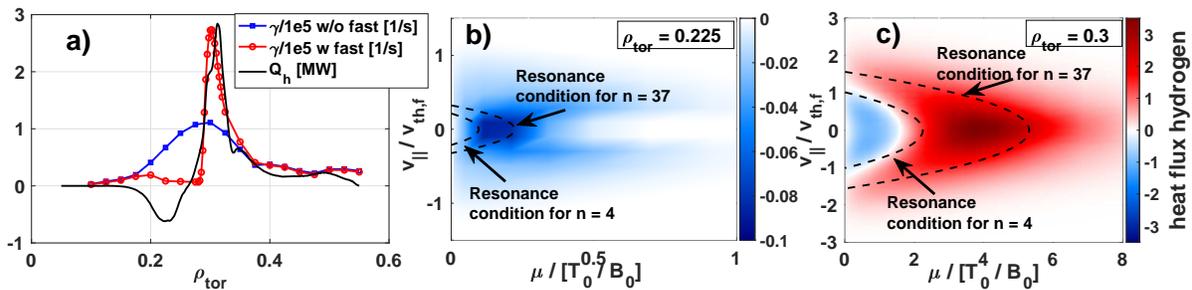}
\par\end{center}
\caption{a) Linear growth rates at the toroidal-mode number $n = 21$ - one of the dominant modes in the nonlinear simulations - with (red) or without (blue) hydrogen minority. The black line denotes the radial profile of the energetic particle heat flux averaged over $t [c_s / a] = [400 - 600]$. Velocity space structure of the saturated nonlinear heat flux (in gyroBohm units) averaged over each directions at b) $\rho_{\rm tor} = 0.225$ and c) $\rho_{\rm tor} = 0.3$. The black contour lines indicate the phase-space resonance positions for $n = 4$ and $n = 37$ at $z = 0$. Electromagnetic fluctuations are neglected in b) and c).}
\label{fig:fig5}
\end{figure*}
Interestingly, Fig.~\ref{fig:fig4}b reveals that the localized turbulence suppression, characteristic of the F-ATB, is largely observed also by neglecting the electromagnetic fluctuations, supporting the nature of this basically electrostatic trigger mechanism. We note that these findings cannot be explained in terms of rational surfaces in the safety factor \cite{Joffrin_NF_2003} (being the magnetic geometry fixed in the simulations with/without fast particles) or by large-scale electromagnetic activity triggered by the supra-thermal particles at the rational surfaces \cite{Joffrin_NF_2003,Wong_NF_2004} (given the electrostatic nature of the F-ATB).

{\em F-ATB trigger mechanism.} The fast ion physical mechanism responsible for the F-ATB generation is identified as a resonant interaction between ion-driven turbulence and supra-thermal particles \cite{DiSiena_NF2018,DiSiena_PoP2019,W7X}. An energy redistribution occurs between fast particles and plasma turbulence when the fast ion magnetic-drift frequency (in GENE normalized units) $\omega_{d,f} = -k_y T_f E_f ({\bf B}_0 \times {\bf \nabla}B_0)\cdot \hat{y}/(q_f B_0^3)$ is close to the frequency of the underlying ITG micro-instability $\omega_r$. This resonance condition is controlled by the energetic particle equivalent temperature $T_f$. The quantity $q_f$ represents the hydrogen charge (normalized to the bulk deuterium one), $k_y$ the bi-normal mode number, while $E_f = (2v_\shortparallel^2 + \mu B_0)$.

The direction of this resonant energy exchange is determined by the radial derivative of the supra-thermal ion distribution function. As shown with analytic theory and nonlinear flux-tube simulations ~\cite{DiSiena_NF2018}, an effective turbulence suppression is achieved when (i) the fast particle logarithmic temperature gradient overcomes the corresponding density one and (ii) the resonance condition $\omega_{d,f} = \omega_r$ is satisfied in the negative fast ion drive region for those modes which contribute most to the transport. When these two constraints are simultaneously fulfilled, a stabilizing energetic particle effect on the ITG thermal drive is maximized, leading to a turbulence suppression. The resonance condition can also be satisfied in a velocity space region where the fast particle drive is positive \cite{DiSiena_PoP2019}. In such cases, the wave-particle resonance amplifies a positive (destabilizing) contribution of the supra-thermal particles, resulting in a destabilization of the ion-driven instability. This interplay between ITG microinstability and the energetic particles is well observed also in nonlinear studies. Inward (outward) supra-thermal ion particle and heat fluxes are observed in correspondence with the strongest fast ion stabilization (destabilization). Therefore, stabilizing and destabilizing energetic particle effects on plasma turbulence occur at different radial positions - depending on the local values of the fast particle parameters.

These predictions are consistent with the GENE results. As illustrated in Fig.~\ref{fig:fig5}a, the dominant local linear growth rate of the $n = 21$ toroidal-mode exhibits an almost full suppression in the radial region $\rho_{\rm tor} = [0.2, 0.25]$. In this radial domain, the wave-fast particle resonance interaction is predicted to be more effective, and the hydrogen minority contribution to the most unstable ITG mode is dominated by the stabilizing region of velocity space. Due to the local changes in the fast ion temperature and density profiles, the effect of supra-thermal particles on plasma turbulence turns from stabilizing to destabilizing in $\rho_{\rm tor} = [0.25, 0.3]$. The ITG growth rate largely increases compared to the reference case without the hydrogen minority. No significant difference is observed by selecting a different toroidal mode numbers.

These flux-tube linear findings are in agreement with the global electromagnetic nonlinear GENE simulations. More precisely, the time-averaged heat flux profile of the hydrogen minority goes from a negative (inward) flux - in the region the fast particle stabilize the linear ITG growth rates - to positive (outward) - where the ITG growth rates are largely destabilized. This is shown in Fig.~\ref{fig:fig5}a (which zooms on the black curve of Fig.~\ref{fig:fig2}c).

This physical interpretation is further supported by the results of Fig.~\ref{fig:fig5}b-c, where the velocity space structure of the energetic particle heat flux $Q_h$ is illustrated at different $\rho_{\rm tor}$. In particular, $Q_h$ exhibits only negative values in Fig.~\ref{fig:fig5}b at the radial position where the largest stabilizing effect is observed in Fig.~\ref{fig:fig5}a ($\rho_{\rm tor} = 0.225$). Interestingly, this beneficial region is localized in phase-space where the resonance condition of the most relevant modes is matched (area delimited by the dotted black lines in Fig.~\ref{fig:fig5}b). Therefore, at $\rho_{\rm tor} = 0.225$, the resonance interaction strongly affects the shape of $Q_h$ maximizing a stabilizing fast ion contribution, leading to inward fast particle fluxes and to a significant ITG stabilization.
On the contrary, the wave-particle resonance interaction enhances the turbulence drive at $\rho_{\rm tor} = 0.3$, as shown in Fig.~\ref{fig:fig5}c, where a predominantly positive velocity space structure of the fast ion heat flux is observed. The largest heat flux contribution lies again within the area delimited by the resonance conditions of the modes that drive most of the turbulent transport. Interestingly, also at these radial positions a suppression of the heat fluxes associated to the bulk species is observed, as shown in Fig.~\ref{fig:fig2}c.

{\em Conclusions.} This Letter presents direct evidence of a predicted-first energetic particle triggered turbulent transport barrier. The existence of this new type of ITB is demonstrated via global gyrokinetic simulations with realistic ion-to-electron mass ratio, collisions, and fast ions modeled with realistic background distributions. This is the first time a gyrokinetic code captures the formation of an ITB retaining all of these physical effects at once.

These findings are based on a realistic ASDEX Upgrade plasma discharge carefully optimized to maximized the beneficial role of supra-thermal particles predicted theoretically.

We provide a physics understanding of this effect, demonstrating that the trigger mechanism responsible for the generation of the F-ATB is a resonance interaction between supra-thermal particles - generated via ICRH - and ITG micro-turbulence, whose overall effect is the formation of localized layers in the $E \times B$ velocity and thus to a transport barrier.

The results within this Letter may be applied regularly in magnetic confinement experiments with significant ICRF heating, such as envisioned for SPARC \cite{SPARK_2020}, to access new types of high-performance discharges with reduced transport levels and enhanced confinement.

This work has been carried out within the framework of the EUROfusion Consortium and has received funding from the Euratom research and training programme 2014-2018 and 2019-2020 under grant agreement No 633053. The views and opinions expressed herein do not necessarily reflect those of the European Commission. The simulations presented in this work were performed at the Cobra HPC system at the Max Planck Computing and Data Facility (MPCDF), Germany. Furthermore, we acknowledge the CINECA award under the ISCRA initiative, for the availability of high performance computing resources and support. The authors would like to thank G. Merlo, R. Fischer, P. David, R. McDermott, G. Vogel, T. Puetterich and G. Tardini.

\end{document}